\documentclass[aps,twocolumn,groupedaddress,amsmath,ams-fonts,superscriptaddress]{revtex4-1}

\usepackage[pdftex]{hyperref}   %for pdftex
\usepackage{graphicx}
\usepackage{dcolumn}% Align table columns on decimal pointB
\usepackage{bm}% bold math
\usepackage{amssymb}
\usepackage{txfonts}
\usepackage{braket}
\usepackage{color}
\def\(#1\){\begin{itemize}#1\end{itemize}}
\newcommand{\bka}{{\bm \kappa}}

\newcommand{\bk}{{\bf k}}
\newcommand{\br}{{\bf r}}
\newcommand{\bq}{{\bf q}}
\newcommand{\epn}{\omega_n}
\newcommand{\epf}{\epsilon_F}
\newcommand{\mG}{\mathcal G }
\def\m[#1\m]{\begin{pmatrix}#1\end{pmatrix}}

\begin{document}
\title{{Generation and control of noncollinear magnetism by supercurrent} } 
\author{Rina Takashima}
\email{takashima@scphys.kyoto-u.ac.jp}
\affiliation{Department of Physics, Kyoto University, Kyoto 606-8502, Japan}
\author{Yasuyuki Kato}
\affiliation{Department of of Applied Physics, The University of Tokyo, Tokyo 113-8656, Japan }
\author{Youichi Yanase}
\affiliation{Department of Physics, Kyoto University, Kyoto 606-8502, Japan}
\author{Yukitoshi Motome}
\affiliation{Department of of Applied Physics, The University of Tokyo, Tokyo 113-8656, Japan }
\date{\today}
\begin{abstract}
When superconductivity couples with noncollinear spin textures, rich physics arises; for instance,  
singlet Cooper pairs can be converted to triplet pairs, and topological superconductors can be realized. For their applications, the controllability of noncollinear magnetism is a crucial issue. 
Here, we propose that a supercurrent can induce and control noncollinear magnetic orders in a correlated metal on top of a singlet superconductor.  
We show that the magnetic instability in the correlated metal is enhanced by the proximity effect of supercurrents, 
which leads to phase transitions from a paramagnetic state to noncollinear magnetic phases with helical or vortexlike spin textures. 
Furthermore, these magnetic orders can be switched by the direction of the supercurrent. 
We also discuss the effect of the Rashba spin-orbit coupling and the experimental realization.   
\end{abstract} 
\maketitle
{\it Introduction.}
The proximity effect of superconductivity, inducing Cooper pairs in non-superconducting materials, has renewed interest for application ranging from spintronics devices to topological superconductivity. In spintronics, spin-triplet Cooper pairs are promising spin 
carriers for efficient spintronics devices with suppressed Joule heating~\cite{Linder2015, Eschrig2015}. 
The interplay between triplet pairs and magnetic moments show rich physics such as a novel type of domain wall dynamics~\cite{Kastening2006,  Brydon2008, Linder2012,Takashima2017}.    
Furthermore, an artificially engineered system with proximity-induced superconductivity is one of the most promising platforms for realizing topological superconductors~\cite{Alicea2010, Sau2010, Mourik2012, Das2012, Nadj-perge2014}. They can host Majorana fermions, whose non-Abelian statistics might be used for topological quantum computing~\cite{Nayak2008}. 
 
In these applications of superconductivity, noncollinear magnetic orders play essential roles. Triplet Cooper pairs can be generated from a singlet superconductor via the coupling to noncollinear magnetic moments: noncollinearity of magnetic moments breaks the spin rotational symmetry 
of electrons and can convert singlet Cooper pairs to triplet pairs~\cite{Bergeret2005}.  This is experimentally observed~\cite{Keizer2006, Khaire2010, Robinson2010, Anwar2010}, e.g., in a multilayer system of a conical magnet holmium and a singlet superconductor~\cite{Robinson2010}. 
Moreover, we can engineer a topological superconductor by using noncollinear magnetic moments. One-dimensional $p$-wave superconductivity can be realized using {a} helical spin texture~\cite{Choy2011, Nadj-Perge2013, Klinovaja2013, Vazifeh2013, Braunecker2013}, and two-dimensional $p+ip$-wave superconductivity can be realized using a {magnetic} skyrmion texture~\cite{Nakosai2013, Chen2015, Li2016}. These schemes do not require the strong relativistic spin-orbit coupling, and hence expand the range of candidate materials. 

Given the interplay between noncollinear spin textures and superconductivity, the tunability of noncollinear magnetism is crucial to externally control the resulting physics. For example, we could turn on or optimize the singlet-triplet conversion, and it might also be possible to manipulate Majorana zero modes by magnetic states. In normal 
states, several ways are known to induce or switch magnetic textures, e.g., applying an electric field~\cite{Ohno2000, Lottermoser2004} or optical pulses~\cite{Stanciu2007, Radu2011}. 
%{In the  coexisting phase of superconductivity and magnetism, the domains of collinear magnetic moments have been studied~\cite{Bulaevskii1983, Bulaevskii1985}. It is shown that a supercurrent can change  the direction of  domains. 
%However, it has been unclear how to induce and control noncollinear magnetism  in the presence of superconductivity. }
{In a superconducting state, it was %also 
shown that a supercurrent can change the direction of magnetic domains~\cite{Bulaevskii1983, Bulaevskii1985}. However, as it is limited to a coexisting phase of superconductivity and magnetism, it is highly desired to establish a versatile way not only to control but also generate noncollinear magnetism by a supercurrent.}

\begin{figure}[b]
\includegraphics[width=4.cm]{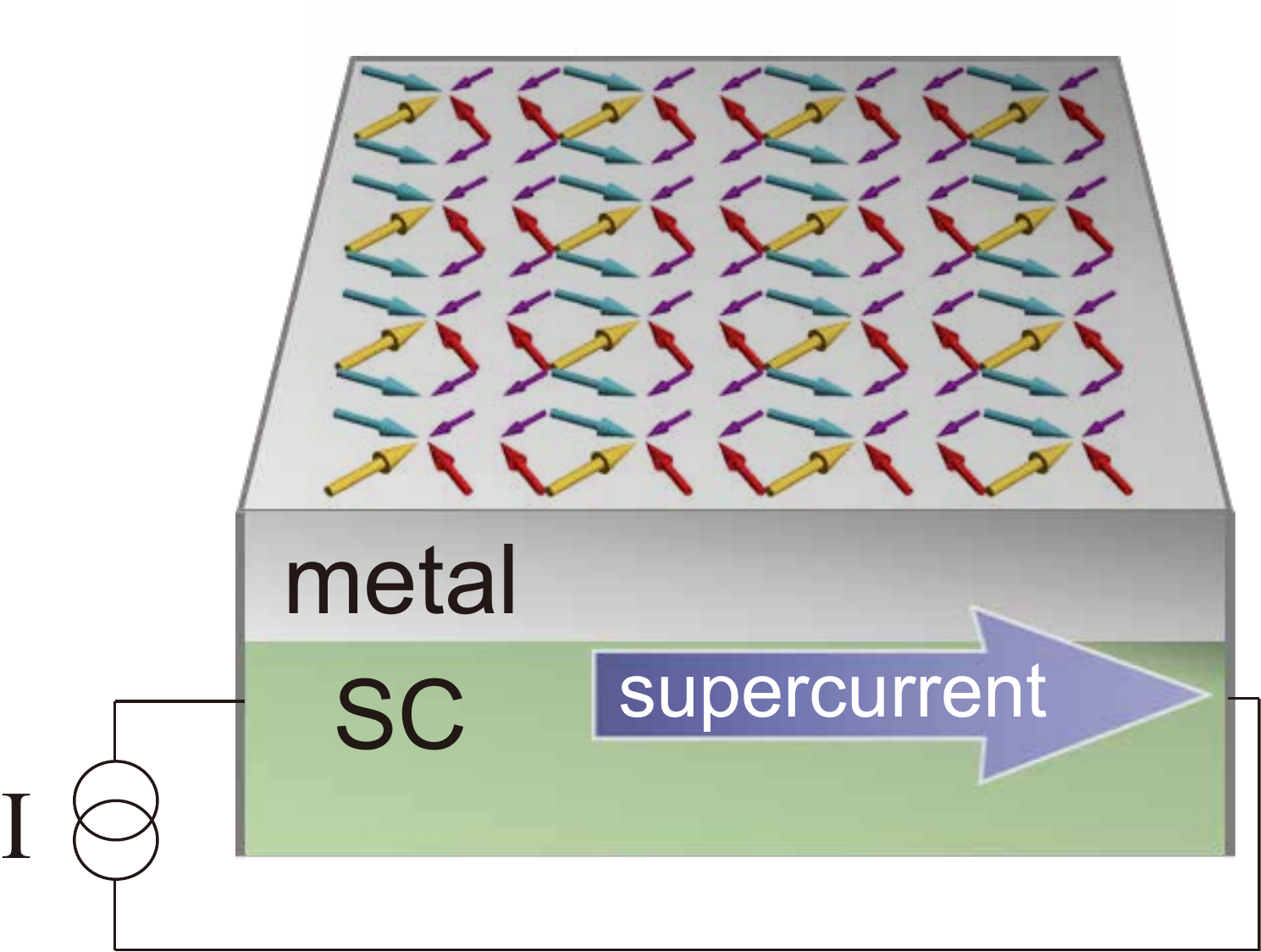} %\label{setup}
\caption{
The proposed setup for an experimental realization.  A metallic layer is  
deposited on  
a bulk singlet superconductor (SC), where a supercurrent is supplied from a current battery. 
}
\label{setup}
\end{figure}

In this Rapid Communication, we propose a way to {induce and control} noncollinear magnetic order by utilizing a supercurrent, in a correlated metal 
in proximity to a singlet superconductor. We consider a { quasi-two-dimensional} metallic layer on top of a singlet superconductor with a supercurrent flow (Fig.~\ref{setup}).  
We demonstrate that a supercurrent induces noncollinear magnetic orders, such as helical (single-$Q$) 
and vortexlike (double-$Q$) orders in a controlled manner. Furthermore, these states can be switched by the direction of the supercurrent. 
Our results open up 
the possibilities to use superconducting correlations as a ``harness" for spin textures. 

{\it Model.}
We consider a {quasi-two-dimensional 
correlated metal} deposited on a bulk singlet superconductor with a supercurrent flow.   A possible experimental setup is shown in Fig.~\ref{setup}. %This can be described 
{The situation can be modeled} by the Hamiltonian $H =H_0+H_{U}$, where we define
\begin{eqnarray}  
&H_0&= \sum_{ \bk \sigma }\xi_{\bk}  c^\dag_{\bk \sigma}c_{\bk \sigma}+\sum_{ij}(\Delta_{ ij} e^{ i\bka \cdot (\br_i +\br_j) }  c_{i \uparrow}^{\dag}c_{j \downarrow}^{\dag}+\text{H.c.}),\\
&H_{U}&=U\sum_i n_{i \uparrow} n_{i \downarrow}.
\end{eqnarray}  
Here $c^\dag_{i\sigma} (c_{i\sigma})$ is a creation (annihilation) operator of itinerant electrons at {site $\br_i =(x_i, y_i)$}  with spin $\sigma$,  $c^\dag_{\bk \sigma} (c_{\bk \sigma})$ is the Fourier transform of $c^\dag_{i\sigma} (c_{i\sigma})$, 
and $\xi_{\bk}$ is the 
energy dispersion measured from the chemical potential. The itinerant electrons have the repulsive Hubbard interaction with the strength $U>0$, and $n_{i \sigma}=c^\dag_{i\sigma} c_{i\sigma}$ is the electron density {operator} of spin $\sigma$. 
The proximity effect to the superconductor is taken into account by $\Delta_{ij}$ and $\bka$; $\Delta_{ij}$ describes a singlet pairing potential between site{s} $i$ and $j$, and $\bka$, a spatial gradient of the superconducting phase, originates from a supercurrent flow. The supercurrent is assumed to be a unidirectional flow, and the supercurrent density {${\bf j}^{\rm sc}$} is 
  proportional to $\bka$ 
  for  $\xi|\bm \kappa|  \ll 1$, where $\xi$ is the coherence length of the bulk superconductor.
 In the following, we study this model within the mean-field approximation as $H \approx H_{0} +H^{\rm MF}_{U}$, wher{e}  
\begin{eqnarray}  
H_U^{\rm MF} =
-\frac{4U}{3}\sum_i {\bf m}_i \cdot {\bf s}_i +\frac{2U}{3}\sum_i |{\bf m}_i|^2.
\label{mf}
\end{eqnarray}  
Here {${\bf  s}_i= \frac{1}{2} \sum_{\sigma_1, \sigma_2} 
c^\dag_{i \sigma_1} \bm \sigma_{\sigma_1 \sigma_2 } c_{i  \sigma_2 }$} 
is the spin density operator at site $i$ with the Pauli matrices  $\bm \sigma =(\sigma^x, \sigma^y,\sigma^z)$ , and  ${\bf m}_i =\langle {\bf s}_i \rangle $ is %a 
{the} mea{n 
f}ield of the spin density. 

{\it Magnetic instability.}
{L}et us first demonstrate that a supercurrent ${\bf j}^{\rm {sc}} \propto \bm \kappa$ enhances an instability toward a noncollinear magnetic order.  From now on, we will focus on an $s$-wave pairing ($\Delta_{ij}=\frac{1}{2}\Delta \delta_{ij}$) for simplicity{, while we will discuss other symmetries later}. 
We calculate the energy functional of ${\bf m}_i$, which is obtained  by integrating out the electron operators:    
\begin{eqnarray}  {\mathcal{E}}[\{ {\bf m}_\bq \}] =\frac{2U}{3}\sum_{\bq}\left(1 - \frac{2U}{3} \chi (\bq) \right)  |{\bf m}_{\bq}|^2, \label{effH} \end{eqnarray}  
where ${\bf m}_{\bq}$ is the Fourier transform of ${\bf m}_i$, and we have taken the lowest order of ${\bf m}_{\bq}$. $\chi(\bq) >0$ is the bare spin susceptibility under a supercurrent~\footnote{See Supplemental Material at xxx for the details of $f(x,y)$ and $g(x,y)$. }.  
The largest peak of $\chi(\bq)$ indicate{s} an instability toward a magnetic order given by the corresponding mode $\bf m_{\bq}$. 

\begin{figure}
\includegraphics[width=8cm]{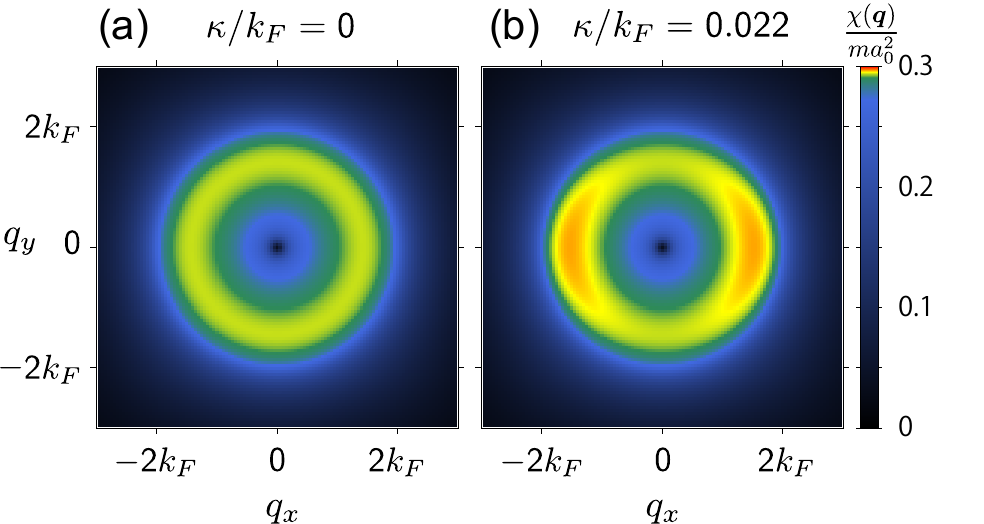}  
\caption{Bare spin susceptibility $\chi(\bq)$ in the low density limit 
for  $|\Delta|/\epf=0.05$ %{
and $k_B T/|\Delta|=1/40 \ll 1$ %. 
{(a) %$\chi(\bm q)$ 
w}ithout a supercurrent {($\bm\kappa=\bm 0$) %. 
and (b)  
w}ith the supercurrent $ \bm \kappa /k_F =(0.022, 0)$.   
 }
\label{fig:chiq} 
\end{figure}

Here we show how a supercurrent changes the profile of $\chi(\bq)$ considering  the low density limit in two dimensions 
at low temperature ($k_{B} T\ll |\Delta|$), where $k_{ B}$ is the Boltzmann constant and $T$ is the temperature. 
We define $\xi_{\bk} = \frac{ k^2}{2m} -\epf$, where $m$ is the electron mass, $\epf \geq 0$ is the Fermi energy, and we set {the reduced Planck constant} $\hbar=1$.  
In a normal state ($\Delta=0$), $\chi(\bq)$ is constant and largest for $ |\bq| \leq  2 k_F$ with $k_F =  \sqrt{ 2m \epf}$, i.e., all modes of $\bf m_{\bq}$  with $|\bq|<2 k_F$ energetically degenerate in Eq.~\eqref{effH}. 
With an $s$-wave singlet correlation without a supercurrent ($\Delta\neq0,\  \bka =\bm 0$), $\chi(\bq)$ is suppressed around  $\bq = \bm 0$ because of the spin gap associated with the superconductivity [Fig.~\ref{fig:chiq}(a)]. This is called the  
Anderson-Suhl mechanism~\cite{Anderson1959}. 
Now, $\chi(\bq)$ is further deformed {from the ring structure} by applying a supercurrent, which is given by ${\bf j}^{\rm {sc}} = -e\frac{n_{\rm sf}}{m} \bka$ with the electron charge $-e$ and the superfluid density $n_{\rm sf}$. 
A supercurrent introduces peak structures around $\bq^* \sim \pm 2 k_F \hat{\bm \kappa} $ 
[Fig.~\ref{fig:chiq}(b)], with $\hat{\bka}=\bka/|\bm \kappa|$.

Such an increase induced by a supercurrent can be analytically obtained for a small current density. 
For $k_{B}T \ll |\Delta|$,  we can expand $\chi(\bq)$ to the second order of $\bm \kappa$ as  
\begin{eqnarray}  \chi(\bq)-\chi_{\bm \kappa=\bm 0}(\bq) %\nonumber\\
  %&
  &=
\frac{ a_0^2|\bm \kappa|^2}{\epf} f \left( \frac{q}{k_F},\frac{|\Delta|}{\epf} \right)%\nonumber \\
+\frac{a_0^2\left(\bm \kappa \cdot \hat{\bq}\right)^2}{\epf } g\left( \frac{q}{k_F},\frac{|\Delta|}{\epf} \right) 
,  \label{chi}\end{eqnarray}   
 where $\chi_{\bm \kappa =\bm 0}(\bq)$ is the bare spin susceptibility at $\bm \kappa =\bm 0$, $a_0$ stands for the inverse of the momentum cutoff, $f(x,y)$ and $g(x,y)$ are the dimensionless functions, and we define $q=|\bq|$ and $\hat{\bq}=\bq/q$. 
The directional dependence arises from the second term in the right-hand side, and $g\left( \frac{q}{k_F},\frac{|\Delta|}{\epf} \right) $ is proved to be positive and has a peak around $q\sim 2k_F$.  
We also note that, for $|\Delta|/\epf \ll1$, we obtain $|g(q/k_F,|\Delta|/\epf )| \gg |f(q/k_F,|\Delta|/\epf)|$~[32]. 
As a result, $\chi(\bq)$ increases around $\bq^{{*}} \sim \pm 2k_F\hat{\bka}$ by a supercurrent, and the magnetic instability with the particular wave number $\bq^*$ is selected out of the degenerate ring. 
{The enhancement of the spin susceptibility by a supercurrent depends on the form of the Fermi surface; it can be larger depending on systems.
}

{\it Magnetic phase diagrams.}
On a lattice with a higher density of electrons, the bare spin susceptibility %becomes anisotropic 
{$\chi(\bq)$ is no longer isotropic: it} %and  
generally has peak structures  even without a supercurrent{, reflecting the lattice structure and the energy dispersion}. Therefore, magnetic instabilities %result 
{arise} from the interplay of the %lattice geometry 
{anisotropy of $\chi(\bq)$} and the current direction. 

In the following, we investigate magnetic {ground-state} phase diagrams for %the %ground states by numerical diagonalization
{a} {lattice mode{l} %s 
numerically}.     
We consider a square-lattice model with the energy dispersion $\xi_{\bk}= -2t(\cos (k_x a) +\cos(k_y a)) -\mu$, where $t$ denotes the hopping amplitude between the nearest-neighbor sites, $a$ is the lattice constant, and $\mu$ is the chemical potential. 
In the calculations, we set $t=0.5$, %\  
$\mu= -1.48$, %\ 
{and} $\Delta=0.05$. With these parameters, $\chi_{\bka =\bm 0}(\bq)$ has four equivalent peaks at $\bq=(q_x, q_y) =(\pm Q^*,0), \ (0, \pm Q^*)$ with $Q^* \simeq 2\pi/3a$.   
%{In an applied supercurrent, these peaks can be modulated independently, depending on the current direction.} 

\begin{figure}[t!]
\includegraphics[width=\columnwidth]{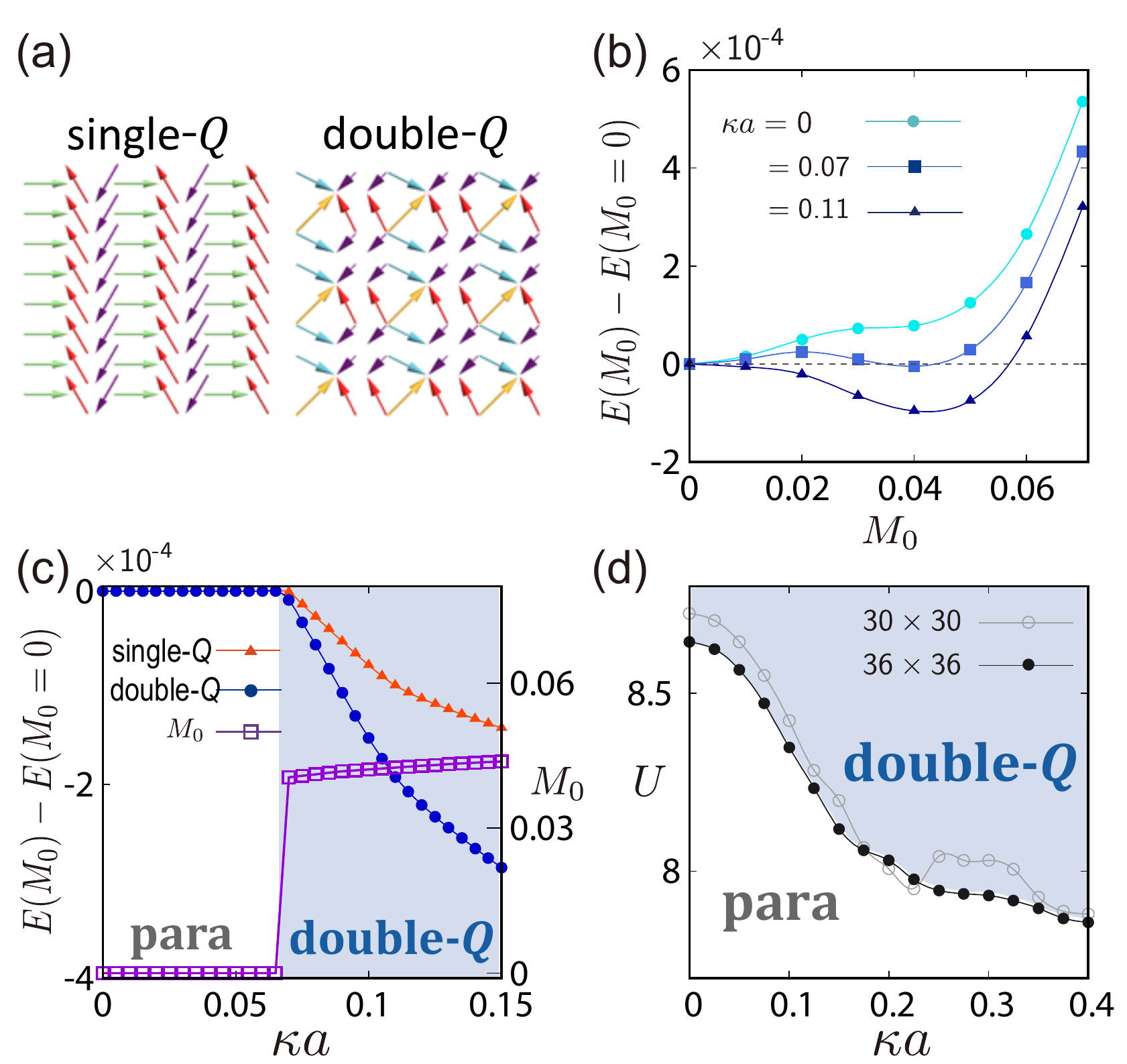}
\caption{
Magnetic ordering induced by a supercurrent given by $\bka=\frac{\kappa}{\sqrt 2}(1,1)$. 
(a) Schematic pictures of a single-{$Q$} state (left) and a double-{$Q$} state (right). 
(b) Energy of the double-{$Q$} state as a function of $M_0$  for different $\kappa$ at $U=8.58$. 
(c) Energies of the single-{$Q$} and double-{$Q$} state{s} and the magnetization of the double-{$Q$} state as functions of $\kappa$ at $U=8.58$. 
(d) $\kappa a$-$U$ phase diagram for the ground state. Phase boundaries obtained in different system sizes are plotted. 
%{We note that $ E(M_0)$ is the energy of the mean-field Hamiltonian ($H_0+H_U^{\rm MF}$) with the optimized $M_0$.  
%The energy unit is taken as $2t =1$. }  
}
\label{fig:phase}
\end{figure}

To {elucidate the ground-state phase diagram in the presence of the supercurrent}, we perform variational calculations.  
Referring to the previous %{
studies~\cite{Agterberg2000a,  Hayami2014, Ozawa2016}%}
, %magnetic phases in the Kondo lattice 
we %consider 
{assume} tw{o %simple 
v}ariational ansatz for magnetic configurations: helical (single-{$Q$}) and vortex{like} (double-{$Q$}) states, which are described by 
\begin{eqnarray}  
{\bf  m}_i^{{1Q}} %{\rm single} &
= M_0 \begin{pmatrix}
 \cos(Q  x_i ) \\
\sin(Q x_i)\\ 
0
\end{pmatrix},
\hspace{10 pt}
{\mathbf m}_i^{{2Q}} 
= M_0 \begin{pmatrix}
 \cos(Q  x_i ) \\
\cos(Q y_i)\\ 
0
\end{pmatrix}, \label{ans2}
\end{eqnarray}  
respectively 
[Fig.~\ref{fig:phase}(a)].
Here $M_0$ is the amplitude of the magnetization and $Q$ is the wave number of modulation, and they are the variational parameters~\footnote{We consider $Q$ at and close to the peak of $\chi(\bq)$.}. 
{We note that the spin-rotational symmetry exists in the absence of the spin-orbit coupling, and hence rotations of all the spins does not change the energy (the effect of the spin-orbit coupling will be discussed later).}
{The} single-{$Q$} state is characterized by the single mode, ${\bf m}_{\bq=(\pm Q, 0)}$, wheres {the} double{-$Q$} state consists of the two modes: ${\bf m}_{\bq=(\pm Q, 0)}$ and ${\bf m}_{\bq=(0, \pm Q)}$
~\footnote{We note that the stability of the double-{$Q$} state might be underestimated with the above simple ansatz, as a noncoplanar component is neglected~\cite{Ozawa2016} }.
{By substituting Eq.~(\ref{ans2}) for ${\bf m}_i$, we numerically diagonalize the mean-field Hamiltonian 
$H_{0}+H_{U}^{\rm MF}$ 
for finite-size systems with the open boundary conditions, and %obtain 
optimize the ground state energy $E(M_0)$ with respect to $M_0$ and $Q$}. The results for $30 \times 30$ sites are shown in Figs.~3 and 4, wheres Fig.~3(d) also shows the result for $36 \times 36$ sites. {We set the energy unit as $2t=1$.}

First, we consider a supercurrent given by $\bm \kappa =(\kappa_x,\kappa_y) =\frac{\kappa}{\sqrt 2} (1, 1)$. 
In this case, the four peaks in $\chi(\bq)$ {at $\bq=(\pm Q^*,0)$ and $(0, \pm Q^*)$} are equally enhanced.  
In Fig.~\ref{fig:phase}(b), we show the $M_0$ dependence of the energies of the double-$Q$ state for several values of $\kappa a$, where the energies are measured from that with $M_0=0$. We note that the energy for the single-$Q$ state is always higher than that for the double-$Q$ state. 
For small $\kappa$, the system is in the paramagnetic state ($M_0 =0$).  
With $\kappa$ increased, the first{-}order phase transition to the double-$Q$ ordered state occurs, and the amplitude of the magnetization $M_0$ shows a jump at the transition. Figure~\ref{fig:phase}(c) shows the $\kappa$ dependences of the system energies for the single-{$Q$} and double-{$Q$} ordered states and $M_0$ in the double-$Q$ state. For $\kappa$ larger than the critical value, the double-{$Q$} ordered state has the lower energy than the paramagnetic state as well as the single-$Q$ state. Therefore, the supercurrent induces a magnetic order transition from the paramagnetic state to the double-{$Q$} state. 
The critical $\kappa$ is summarized in the ground-state phase diagram in Fig.~\ref{fig:phase}(d).  
The result shows that a larger supercurrent induces the magnetic instability at a smaller interaction strength $U$.
We note that, for $\kappa a \gtrsim0.15 $, the jump of $M_0$ becomes small (not shown), suggesting that the phase transition can possibly be continuous in the large $\kappa$ region.  The features found here are confirmed at different system sizes, and the phase boundary becomes smoother for the larger system [Fig.~\ref{fig:phase}(d)]. 

{
%The energy gain by the double-$Q$ ordering is in the order of $10^{-4}t$ as shown in Figs.~3(b) and (c). 
In a realistic situation with finite thickness, a finite-temperature phase transition would occur. The critical temperature $T_c$ of the magnetic ordering would be roughly given by $\sim U M_0$,  which is on the order of $0.1 t$ within the mean-field approximation [see Fig.~3(b)]. $T_c$ will be suppressed by fluctuations, but we expect it could  remain at a finite temperature. 
% the observable range of temperature.
}

\begin{figure}[t]
\includegraphics[width=\columnwidth]{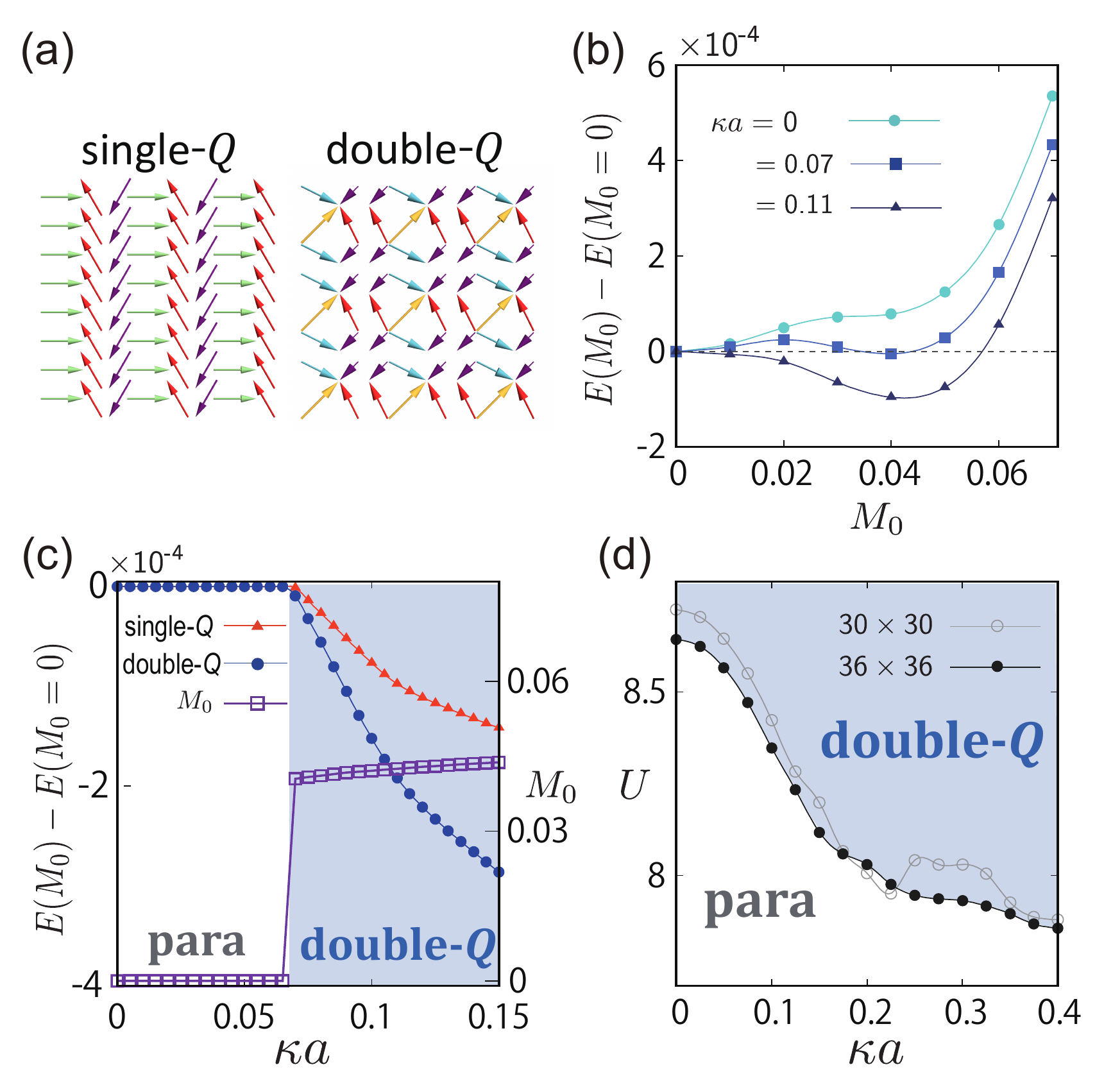}
\caption{
Switching of magnetic orders by the supercurrent %angle variation $\delta \theta$ 
{direction}.  (a) Energies of the single-{$Q$} and double-{$Q$} states as functions of {the angle} $\delta \theta$ at $U=8.49$. 
(b) %$U$-$\delta \theta$ 
{$\delta\theta$-$U$} phase diagram for the ground state. 
 }
\label{fig:angle}
\end{figure}

Furthermore, the stable magnetic state can be switched by changing the supercurrent direction, as the peaks of $\chi(\bq)$ can be modulated independently depending on the current direction. For example, once a current is applied along the $x$ direction, the instability of ${\bf m}_{\bq=(\pm Q,0)}$ is enhanced compared to ${\bf m}_{\bq=(0,\pm Q)}$.
To demonstrate how the stable magnetic state changes, let us rotate the current direction from $(1,1)$ to $(1,0)$ by defining $\bm \kappa = \kappa (\cos (45^\circ-\delta\theta), \sin (45^\circ-\delta\theta))$ [the inset of Fig.~\ref{fig:angle}(a)]. 
 Figure~\ref{fig:angle}(a) shows the energies for the single-{$Q$} and double-{$Q$} states as functions of the current angle {$\delta\theta$}.
While increasing $\delta\theta$, the double-$Q$ order switches to a single-{$Q$} order, which remains stable up to 
$\delta\theta =45^\circ$, i.e., for the current along the $x$ direction.      
The resulting phase diagram {while changing $U$} is summarized in Fig.~\ref{fig:angle}(b). 

%We consider that an $s$-wave superconductor MgB$_2$ would be a candidate superconductor in the heterostructure  because it has a small coherence length $\xi$, given by $a/\xi \sim 10^{-1}$~\cite{Xu2001}.  
{In order to substantiate the effects found here, we need a superconductor which is robust against the supercurrent. MgB$_2$ would be a prime candidate, as it has a small coherence length $\xi\sim 10a$~\cite{Xu2001}.
At a low temperature, the upper limit of a supercurrent is ideally given by $\kappa /k_F  < (\xi k_F)^{-1}$~\cite{Tinkham1996}. %To reach a larger $\kappa /k_F$, it is desired to use a superconductor with a higher upper critical field $\propto \xi^{-2}$. 
In  MgB$_2$,  $\kappa /k_F$ may be of the order of $10^{-1}\sim 1$, which covers the range in Fig.~3(d). %depending on $k_F a$.   
}
%Realistic values of $\kappa /k_F$ depend on the property of the bulk superconductor and {the} correlated metal. 
%At a low temperature, the upper limit is ideally given by $\kappa /k_F  < (\xi k_F)^{-1}$ with $\xi$ being the coherence length of the bulk superconductor~\cite{Tinkham1996}. To reach a larger $\kappa /k_F$, it is desired to use a superconductor with a higher upper critical field $\propto \xi^{-2}$ and a correlated metal with a lower carrier density. For example, the coherence length of the canonical $s$-wave superconductor MgB$_2$ is $a/\xi \sim 10^{-1}$~\cite{Xu2001}, and {hence,} $\kappa /k_F$ may be of the order of $10^{-1}\sim 1$ depending on $k_F a$. 
%The parameter used in Fig.~3(d) is within this range. 
We note that  the above upper limit can be lowered due to the orbital depairing, which could be suppressed by  the geometry of the superconductor to some  extent~\cite{Tinkham1996}.  
{Meanwhile, the deposited metal can be a generic correlated electron system, but one closer to magnetic instability would be better. }

{\it Other gap symmetries.} 
We have discussed an $s$-wave pairing, which has a nodeless isotropic gap in the momentum space.  
Node structures in the superconducting gap render the bare spin susceptibility anisotropic as a lattice does, and the magnetic instability is enhanced particularly for $\bq$ in the directions of the nodes. For example, a $d_{x^2-y^2}$-wave gap has nodes in the $(\pm \pi, \pm \pi)$ direction, and the magnetic modes ${\bf m}_{\bq\sim (\pm {Q}, \pm {Q})}$ have the dominant instability in the absence of supercurrents. 
Thus, the stable magnetic state is further flexibly controlled by the pairing symmetry in addition to  
the current direction, lattice geometry, and energy dispersion. 

{\it Rashba spin-orbit coupling.}
At the interface between a metal and a superconductor, the Rashba spin-orbit coupling may be important due to the broken mirror symmetry along the $z$ axis, the out-of-plane direction. It can be written as
\begin{eqnarray}  
H_{\rm so} = \alpha  \sum_{\bk \sigma_1 \sigma_2} {\bf g}(\bk)\cdot (c^\dag_{\bk \sigma_1} \bm \sigma_{\sigma_1 \sigma_2} c_{\bk \sigma_2}), 
\end{eqnarray}  
where $\alpha$ is the magnitude of the Rashba spin-orbit coupling and ${\bf g}(\bk) =(k_y a_0,-k_x a_0, 0) $ in the continuum limit. 
The spin-orbit coupling breaks the spin rotational symmetry, and Eq.~\eqref{effH} is replaced by  
$E[\{ {\bf m}_\bq \}] =\frac{2U}{3}\sum_{\bq}\left(1 - \frac{2U}{3} \chi^{\mu \nu} (\bq) \right)   m^\mu_{-\bq} m^\nu_{\bq} $. 
Thus, the magnetic instability is dictated by the peak wave number in the anisotropic susceptibility tensor $\chi^{\mu \nu} (\bq)$. 

More importantly, with a supercurrent, the superconducting analog of the Rashba-Edelstein effect arises~\cite{Edelstein1995}. A spin polarization is induced by the supercurrent flow owing to the absence of the mirror symmetry. The first-order correction of $\kappa$ leads to the additional term in the energy functional as 
\begin{eqnarray}  
\delta \mathcal E[\{ {\bf m}_{\bq}\}] =\mathcal K \sum_i (\hat{\bf z}  \times \bm \kappa )\cdot  {\bf m}_{i},
\end{eqnarray}   
where $\mathcal K$ is the odd function of $\alpha$, and $\hat{\bf z}$ is the unit vector in the $z$ direction~[32]. 
This term acts like an in-plane magnetic field, e.g., a supercurrent in the $x$ direction gives an effective magnetic field in the $y$ direction.  This results in the modulation of the noncollinear spin textures. 

Hence, the Rashba spin-orbit coupling locks the spin-spiral plane {by the spin anisotropy}, and moreover, modulates the spin texture by combining with a supercurrent. Even in this case, we believe that our mechanism to control stable magnetic orders remains robust, as $\chi^{\mu \nu} (\bq)$ is modified by a supercurrent, as well as the gap symmetry. It would be interesting to explore the resulting magnetic phase diagrams, and we leave further discussion for future work. 

{\it Summary and discussion.}
To summarize, we have proposed a way to induce and switch noncollinear spin textures by a  supercurrent in the presence of the superconducting proximity effect. Our results can be useful to realize and control physics raised by the interplay between 
superconductivity and noncollinear magnetism, e.g., the singlet-triplet conversion of Cooper pairs and engineering topological superconductivity. 

The mechanism behind is general, as we have shown in the continuum model. 
Using different lattice geometries as well as different pairing symmetries, we can control a broad range of magnetic states. For example, we might be able to create and annihilate Skyrmions by a supercurrent in a triangular lattice, noting that Skyrmion crystals can be realized in the Kondo lattice model, {where conduction electrons couple with localized magnetic moments~\cite{Ozawa2017}.

The previous works in the Kondo lattice model~\cite{Ozawa2016, Ozawa2017} have employed unbiased numerical simulations to show multiple-$Q$ magnetic orders including Skyrmion crystals. Since the Kondo lattice model is related with the mean-field approximation for the Hubbard model discussed in our paper, we deduce that our conclusion remains robust beyond the variational approach.  

In our study, we have assumed that the superconducting proximity effect is robust. %For a self-consistent treatment, one needs to optimize both superconducting and magnetic order parameters layer by layer, and hence it is beyond the scope of this work.  
In reality, there is a feedback from the magnet to the superconductor;
 the interplay should be dealt with a self-consistent calculation of both superconducting and magnetic order parameters layer by layer. This costs much computational cost, and hence, we leave it for future study. 
Nevertheless, we believe that our results already capture the essential physics %that will be obtained in the self-consistent approach 
when the proximity effect from  the bulk superconductor is large and  the bulk superconductivity is stable.  }

We finally note that it would also be interesting to explore what happens with a supercurrent being switched off. % due to the first-order nature of the transition in an $s$-wave case. 
It is important, especially for the application, to clarify whether the magnetic order persists as a metastable state, and how the phase of pairing and magnetic order relax. 

\if0
The mechanism studied in this paper can also be applied to the Kondo lattice 
systems, where conduction electrons couple to localized moments. 
It suggests that not only a heterostructure composed of a heavy-fermion compound and a superconductor but also a heavy-fermion superconductor can lead to further rich control of magnetism, owing to complex magnetic interactions inherent in the heavy-fermion systems.
\fi

We would like to thank R. Ozawa for fruitful discussions. 
This research was supported by Grants-in-Aid for Scientific Research under Grants 
Nos. JP15J01700, JP15H05884, JP16H00991, JP15K05164, JP15H05745, and JP15K05176.
R. T. was supported by a JSPS Fellowship for Young Scientists

%確認用
%takashima　JSPS: 15J01700
%Y. Motome: 15K05176
%Y.Yanase：  JP15H05884, JP16H00991, JP15K05164, and No. JP15H05745 
\bibliographystyle{apsrev4-1}
%\bibliography{C:/Users/r3tak/Documents/Mendeley-files/library}

%merlin.mbs apsrev4-1.bst 2010-07-25 4.21a (PWD, AO, DPC) hacked
%Control: key (0)
%Control: author (72) initials jnrlst
%Control: editor formatted (1) identically to author
%Control: production of article title (-1) disabled
%Control: page (0) single
%Control: year (1) truncated
%Control: production of eprint (0) enabled

%

\begin{widetext}
\newpage
\begin{center}
\vspace{1cm} 
\textbf{{\large Supplemental Material for \\ \vspace{3mm} ``Generation and control of noncollinear magnetism by supercurrent''} }
\end{center}
\setcounter{equation}{0}
\setcounter{figure}{0}
\setcounter{table}{0}
\setcounter{page}{1}
\makeatletter
\def\[#1\]{\begin{align}#1\end{align}}
\renewcommand{\theequation}{S\arabic{equation}}
\renewcommand{\thefigure}{S\arabic{figure}}
\renewcommand{\bibnumfmt}[1]{[S#1]}
\renewcommand{\citenumfont}[1]{S#1}

{
\section{Derivation of  Eq.~(4) }
In this section, we show the derivation of Eq.~(4). 
We start from the Hamiltonian given by Eqs.~(1) and (2) in the main text:
\[
H_0+H_{U}&= \sum_{ \bk \sigma }\xi_{\bk}  c^\dag_{\bk \sigma}c_{\bk \sigma}+\sum_{ij}(\Delta_{ ij} e^{ i\bka \cdot (\br_i +\br_j) }  c_{i \uparrow}^{\dag}c_{j \downarrow}^{\dag}+\text{H.c.})+U\sum_i n_{i \uparrow} n_{i \downarrow}.
\]
Noting that $H_U= \frac{U}{2} \sum_i (n_{i \uparrow} + n_{i \downarrow}  ) -\frac{2 U}{3} \sum_i  {\bf s}_i  \cdot {\bf s}_i $,  
we introduce the Hubbard-Stratonovich fields to decouple the spin-spin interaction.  
The partition function $Z$ is given by

\[
Z=& \int \mathcal D( \psi_{\bk \sigma}, \bar{\psi}_{\bk \sigma}, {\bf m}_i ) \exp\left[-  S(\psi_{\bk \sigma}, \bar{\psi}_{\bk \sigma}, {\bf m}_i )
   \right],\\
S(\psi_{\bk \sigma},  \bar{\psi}_{\bk \sigma},  {\bf m}_i )
=&\int_0^\beta d\tau \left[  \sum_{ \bk \sigma } \bar{\psi}_{\bk \sigma}(\tau)(\partial_\tau + \xi_{\bk} )  \psi_{\bk \sigma}(\tau)+\sum_{ij}(\Delta_{ ij} e^{ i\bka \cdot (\br_i +\br_j) }  \bar{\psi}_{i \uparrow}(\tau) \bar{\psi}_{j \downarrow}(\tau)+\text{H.c.}) \right. \nonumber\\ 
&\left. -\frac{4U}{3}\sum_{i \sigma_1 \sigma_2 } {\bf m}_i(\tau) \cdot \frac{1}{2}\left(\bar{\psi}_{\bk \sigma_1} 
\bm \sigma_{\sigma_1, \sigma_2} {\psi}_{\bk \sigma_2}  \right)  +\frac{2U}{3}\sum_i |{\bf m}_i(\tau)|^2 \right], 
\]
where $\tau$ is the imaginary time, $\beta =(k_B T)^{-1}$, $\bar{\psi}_{\bk \sigma}(\tau)$ and $\psi_{\bk \sigma}(\tau)$ are  Grassmann numbers, and $ {\bf m}_i(\tau)$ is the Hubbard-Stratonovich fields. Then we integrate out $ \bar{\psi}_{\bk \sigma}(\tau)$ and $ \psi_{\bk \sigma}(\tau) $ to obtain the effective action. It is given by       
\[
Z_{\rm eff} & = \int \mathcal D({\bf m}_i ) \exp\left[- S_{\rm eff}( {\bf m}_i ) \right], \label{dist}\\
%Z_{\rm eff} & = \int \mathcal D(\bm m_i ) \exp\left[- \int_0^\beta d\tau L_{\rm eff}[ \bm m_i ] \right], \\
%\int_0^\beta d\tau L_{\rm eff}[ \bm m_i ] &=\frac{2U  }{3 }\sum_{\bq, n}\left(1 - \frac{2U}{3} \chi (\omega_n, \bq) \right)  |{\bf m}_{\omega_n, \bq} |^2 +\cdots 
S_{\rm eff}({\bf m}_i ) &=\frac{2U  }{3 }\sum_{\bq, m}\left(1 - \frac{2U}{3} \chi (\omega_m, \bq) \right)  |{\bf m}_{\omega_m, \bq} |^2 +\cdots,  
\]
where we expand the action in the series of ${\bf m}_{\omega_m, \bq}$. 
Here the dynamical spin susceptibility $ \chi (\omega_m, \bq)$ reads 
\[
\chi(\omega_m,\bq)=-\frac{1}{2N \beta} \sum_{n, \bk } {\rm tr} [ \mathcal G_{\bk+\bq}(\epn+\omega_m)  \mathcal G_{\bk }(\epn)], \]
 where the Green function $ \mG_{\bk }(\epn)$ is the $4\times 4$ matrix given by
\[\mathcal G_\bk(\epn) =
\begin{pmatrix}
\left( i { \epn} -\xi_{\bk+\bka} \right)\mathbb I_{2} 
 & -\Delta (i\sigma^y) \\
-\Delta^*(i\sigma^y)^\dag &
 \left( i \epn +\xi_{-\bk+\bka} \right)\mathbb I_{2} 
 \end{pmatrix}^{-1},
\]
 $\mathbb I_{2}={\rm diag}(1,1) $, $\epn=   (2n+1)\pi/\beta $ is the fermionic Matsubara frequency, and $\omega_m =2 m\pi/\beta$ is the bosonic Matsubara frequency.  

When the fluctuations are negligible, we may apply the saddle-point approximation to the functional integral in Eq.~\eqref{dist}.  
%With the saddle-point solution,  the free energy is given by
%\[
%F-F_0 &= -\frac{1}{\beta} \log Z_{\rm eff}, \\
%&\simeq  \frac{2U  }{3 }\sum_{\bq}\left( 1 - \frac{2U}{3} \chi (\omega_n=0, \bq) \right)  |{\bf m}_{\bq} |^2 +\cdots,   
%\]
%where we subtract the free energy for $U=0$, given by $F_0$, from the original free energy $F$.  
In Eq.~(4) in the main text, we have written the action $\beta^{-1} S_{\rm eff} ({\bf m}_i)$ as the energy functional to minimize, where we assume that the saddle-point solution does not depends on the imaginary time. 
The bare spin susceptibility $\chi(\bq)$ in Eq.~(4) in the main text is given by $\chi(\bq)=\chi(\omega_m=0, \bq)$. 
}

\section{Change in spin susceptibility}
\if0
\[
\chi(\bq)=-\frac{T}{2N} \sum_{n, \bk } {\rm tr} [ \mathcal G_{\bk+\bq}(\epn)  \mathcal G_{\bk }(\epn)], \]
 where the Green function $ \mG_{\bk }(\epn)$ is $4\times 4$ matrix given by
\[\mathcal G_\bk(\epn) =
\begin{pmatrix}
\left( i { \epn} -\xi_{\bk+\bka} \right)\mathbb I_{2} 
 & -\Delta (i\sigma^y) \\
-\Delta^*(i\sigma^y)^\dag &
 \left( i \epn +\xi_{-\bk+\bka} \right)\mathbb I_{2} 
 \end{pmatrix}^{-1},
\]
 $\mathbb I_{2}={\rm diag}(1,1) $, $\epn=  T (2n+1)\pi $ is the Matsubara frequency, and we take $k_B =1$ in the following.  
\fi

A supercurrent changes the profile of $\chi(\bq)$ as discussed in the main text. Figure S1(a) shows the profile of $\chi(\bq \parallel \bka )$ for $\bka =\kappa \hat{\bm x} $, and the increase is the largest at $q_x \sim 2 k_F$ in the continuum model.  
For a small current density, the change induced by a supercurrent can be described by the dimensionless functions $f(x, y)$ and $g(x,y)$ [Eq.~(5) in the main text].
$f(x, y)$ and $g(x,y)$ are given by 
\if0\[
\tilde{f} \left(\frac{q}{k_F},  \frac{|\Delta|}{\epf}  \right)&=
2\int\frac{d^2 \tilde k}{(2\pi)^2} \frac{
\tilde E_{\bk}  \left(|\tilde \Delta|^2 (-2\tilde \xi_{\bk}+\tilde \xi_{\bk+\bq}) -\tilde\xi_\bk^2 \tilde \xi_{\bk +\bq} \right)  
+\tilde E_{\bk+\bq}(\tilde \xi_\bk^3 +|\tilde \Delta|^2 \tilde \xi_{\bk +\bq})
  }{ \tilde E_{\bk}^3\tilde  E_{\bk+\bq}( \tilde E_{\bk}+\tilde  E_{\bk+\bq} )^2  }\\
\tilde g \left(\frac{q}{k_F},  \frac{|\Delta|}{\epf}  \right)&=
\frac{4|\bq|^2}{k_F^2}  \int
\frac{d^2 \tilde k}{(2\pi)^2}
 \frac{\tilde E_\bk \tilde E_{\bk+\bq}-\tilde{|\Delta|}^2 -\tilde \xi_\bk \tilde \xi_{\bk+\bq}}{ \tilde E_\bk \tilde E_{\bk+\bq} (\tilde E_\bk +\tilde E_{\bk+\bq})^3 } 
\]\fi
\[
f(x,y )&=
\frac{1}{2\pi^2} 
 \int^{\frac{2\pi}{ k_F a_0}}_0 \int^{2\pi}_{0}\tilde k d\tilde k  d\theta
\frac{
\sqrt{\tilde\xi_1^2+y^2}
  \left(y^2 (-2\tilde \xi_{1}+\tilde \xi_{2}) -\tilde\xi_1^2 \tilde \xi_{2} \right)  
+\sqrt{\tilde\xi_2^2+y^2}(\tilde \xi_1^3 +y^2 \tilde \xi_2)
  }
{ \sqrt{(\tilde\xi_1^2+y^2)^3(\tilde\xi_2^2+y^2)}(\sqrt{(\tilde\xi_1^2+y^2)} +\sqrt{(\tilde\xi_2^2+y^2)}  )^2 },\\
g(x, y) &= \frac{x^2}{\pi^2} \int^{\frac{2\pi}{ k_F a_0}}_0 \int^{2\pi}_{0} \tilde k d\tilde k  d\theta
\frac{ \sqrt{(\tilde\xi_1^2+y^2)(\tilde\xi_2^2+y^2)}-\tilde\xi_1\tilde\xi_2 -y^2}
{ \sqrt{(\tilde\xi_1^2+y^2)(\tilde\xi_2^2+y^2)}(\sqrt{(\tilde\xi_1^2+y^2)} +\sqrt{(\tilde\xi_2^2+y^2)}  )^3},
\]
where we define $\tilde\xi_1 =\tilde k^2-1$ and $\tilde\xi_2 =\tilde k^2 +2 \tilde k x \cos\theta +x^2 -1$.  
We can prove that $g \left(x, y  \right)$ is always positive noting  the relation $\sqrt{(\tilde\xi_1^2+y^2)(\tilde\xi_2^2+y^2)}>|\tilde\xi_1\tilde\xi_2 + y^2   | $.   

Figure~S1(b) shows the numerical values of the above functions. 
$g \left(\frac{q}{k_F},  \frac{|\Delta|}{\epf}  \right)$, which renders the anisotropic change in $\chi(\bq)$, has a peak at $q\sim 2 k_F$. 
We also note that, around the peak, ${g} \left(\frac{q}{k_F},  \frac{|\Delta|}{\epf}  \right)$ is much larger than  ${f} \left(\frac{q}{k_F},  \frac{|\Delta|}{\epf}  \right)$, which gives the isotropic change.  
%Therefore, $\chi(\bq)$ sharply increases around $\bq^{{*}} \sim \pm 2k_F\hat{\bka}$ by a supercurrent. 
\begin{figure}[h]
\includegraphics[width=15cm]{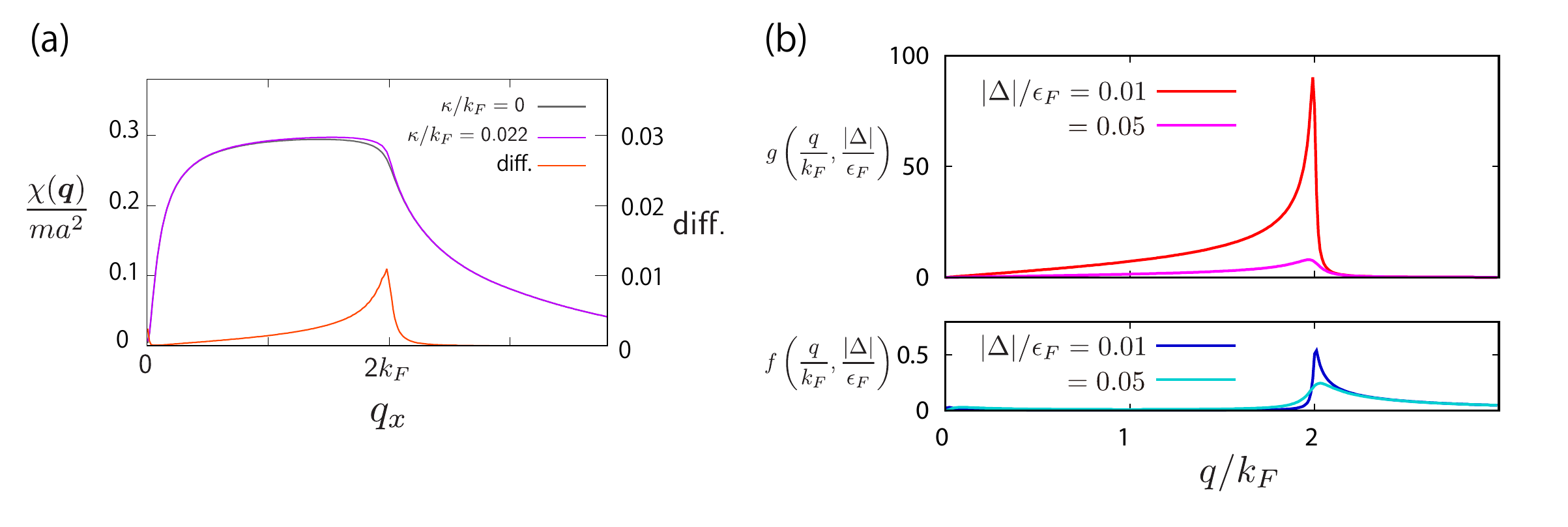}  
\caption{ Increase in $\chi(\bq)$ induced by a supercurrent ($\bka =\kappa \hat{\bm x}$).  
(a) $\chi(\bq\  ||\bka)$ without a supercurrent ($\kappa= 0$)  
and with the supercurrent $ \kappa /k_F =0.022$, and the difference (diff.)  between them.  The difference is largest at $q_x \sim 2k_F$. 
 (b) $g\left( \frac{q}{k_F},\frac{|\Delta|}{\epf} \right) $ and $f\left( \frac{q}{k_F},\frac{|\Delta|}{\epf} \right)$ for different $\Delta/\epf$.  
}
\label{fig:S1} 
\end{figure}　

\section{Rashba spin-orbit coupling}
In this section, we show the derivation of Eq.~(8) in the main text.   
We consider a system with the Rashba spin-orbit coupling given by Eq.~(7) in the main text. Using the Hamiltonian given by $H_0+H^{\rm MF}_U +H_{\rm so}$, we calculate the energy functional of ${\bf m}_i$ by integrating out the electron operators.  
We obtain 
\[
\mathcal E[\{ {\bf m}_\bq \}]  =\frac{2U}{3}  \left(\sum_i m^\mu_{i}\right)\frac{1}{N\beta} \sum_{n, \bk}{\rm tr}[ S^\mu \mG^{\rm so}_\bk(\epn)] +\frac{2U }{3}\sum_\bq  m^\mu_{-\bq} m^\nu_{\bq} \left(
\delta_{\mu\nu} -\frac{2U}{3} \chi^{\mu\nu} (\bq )
\right), \label{efunc}
\]
where we have defined the Green function with the spin-orbit coupling, the spin matrices in the Nambu representation, and the bare spin susceptibility as 
\[
(\mG^{\rm so}_\bk(\epn))^{-1}&= (\mG_\bk(\epn))^{-1} - 
\begin{pmatrix}
\alpha {\bf g}(\bk +\bka) \cdot \bm \sigma &  \\
& -\alpha {\bf g}(-\bk +\bka) \cdot \bm \sigma^{\rm T}
\end{pmatrix},\\
S^\mu &= \begin{pmatrix}
\sigma^\mu & \\
& -(\sigma^\nu)^{\rm T}
\end{pmatrix},\\
\chi^{\mu\nu}(\bq)&=-\frac{1}{2N\beta} \sum_{n, \bk } {\rm tr} [S^\mu \mG^{\rm so}_{\bk+\bq}(\epn) S^\nu \mG^{\rm so}_{\bk }(\epn)].
\]
The bare spin susceptibility is the anisotropic tensor due to the broken spin rotational symmetry. 

Let us focus on the first term in Eq.~\eqref{efunc}. 
We consider the continuum model with $\xi_\bk =\frac{k^2}{2m}-\epf$, ${\bf g}(\bk) =(k_y a_0,-k_x a_0, 0)$, and the supercurrent $\bka =\kappa \hat{\bm x}$  ($\kappa a \ll 1 $).  
We obtain
\[\frac{2U}{3} \frac{1}{N\beta} \sum_{n, \bk}{\rm tr}[ S^\mu \mG^{\rm so}_\bk(\epn)] =\mathcal K \delta_{\mu y}  \kappa ,
\]
to the lowest order of $\kappa$, 
where 
\[
\mathcal K &=- \frac{2U}{3} \frac{1}{N\beta} \sum_{n, \bk}  {\rm tr}[ S^y \mG^0_\bk(\epn) \left( \partial_{\kappa_x} \mG^{0}_\bk(\epn)^{-1}\right)\mG^0_\bk(\epn)],\\
%&= \frac{T}{N} \sum_{n, \bk}  {\rm tr}[ S^x \mG^0_\bk(\epn) \left( \partial_{\kappa_y} \mG^{0}_\bk(\epn)^{-1}\right)\mG^0_\bk(\epn)],
\]
with $\mG^0_\bk(\epn) =\left.\mG^{\rm so}_\bk(\epn) \right|_{\kappa=0}$.
%We note that $\mathcal K$ in Eq.~(8) is given by $\mathcal K=\frac{2UK}{3} $.
$\mathcal K$ is the odd function of $\alpha$, and the explicit form is given by 
\[
\mathcal K&=-\frac{4U  \alpha^3 a_0^5 |\Delta|^2}{3}   %{N}\sum_\bk
\int \frac{d^2k}{(2\pi)^2}
\frac{ k_y^2}{(\xi_\bk^2+|\Delta|^2)^{5/2} },
%&=-\frac{2 \alpha^3}{N}\sum_\bk 
%\frac{ k_x^2 a_0^3 |\Delta|^2}{(\xi_\bk^2+|\Delta|^2)^{5/2} }. 
\]
to the lowest order of $\alpha$ at low temperature $k_BT \ll |\Delta|$.
For the general direction of supercurrents, we have Eq.~(8) in the main text.  

\end{widetext}
\end{document}